\documentclass[a4paper]{jpconf}
\usepackage{graphicx}
\usepackage{subfig}
\usepackage{hyperref}
\begin{document}
\title{Cosmic ray propagation with CRPropa 3}
\author{
  R Alves Batista$^1$,
  M Erdmann$^2$,
  C Evoli$^1$,
  K-H Kampert$^3$,
  D Kuempel$^2$,
  G Mueller$^2$,
  G Sigl$^1$,
  A Van Vliet$^1$,
  D Walz$^2$,
  T Winchen$^3$
}
\address{$^1$ University of Hamburg, II Institut f\"ur Theoretische Physik, Hamburg, Germany}
\address{$^2$ RWTH Aachen University, Physikalisches Institut IIIa, Aachen, Germany}
\address{$^3$ University of Wuppertal, Department of Physics, Wuppertal, Germany}
\ead{walz@physik.rwth-aachen.de}

\begin{abstract}
Solving the question of the origin of ultra-high energy cosmic rays (UHECRs) requires the development of detailed simulation tools in order to interpret the experimental data and draw conclusions on the UHECR universe.
CRPropa is a public Monte Carlo code for the galactic and extragalactic propagation of cosmic ray nuclei above $\sim 10^{17}\,$eV, as well as their photon and neutrino secondaries.
In this contribution the new algorithms and features of CRPropa 3, the next major release, are presented.
CRPropa 3 introduces time-dependent scenarios to include cosmic evolution in the presence of cosmic ray deflections in magnetic fields.
The usage of high resolution magnetic fields is facilitated by shared memory parallelism, modulated fields and fields with heterogeneous resolution.
Galactic propagation is enabled through the implementation of galactic magnetic field models, as well as an efficient forward propagation technique through transformation matrices.
To make use of the large Python ecosystem in astrophysics CRPropa 3 can be steered and extended in Python.
\end{abstract}

\section{Introduction}
The question of the origin of ultra-high energy cosmic rays (UHECRs) continues to be of high interest.
The experimental situation for cosmic rays above $10^{17}\,$eV is promising, with the Telescope Array and the Pierre Auger Observatory covering the northern and southern hemisphere with large exposure.
Recent measurements include the energy spectrum~\cite{Abu-Zayyad:2013qwa,ThePierreAuger:2013eja}, composition~\cite{Abbasi:2014sfa,Aab:2014kda,Aab:2014aea} and anisotropy studies~\cite{Aab:2014ila}.
Interpreting these measurements in terms of concrete astrophysical scenarios requires detailed simulations of cosmic ray propagation from the source to the observer.
In these simulations the deflection of UHECRs need to be computed over several orders of magnitude in energy and length scales, ranging from hundreds of megaparsecs down to galactic kiloparsec scales. Furthermore, all relevant interactions, such as photo-disintegration, pion production and pair production, need to be included.
The simulation tool should be flexible enough to cover the large parameter space of possible astrophysical scenarios, in order to constrain the origin of UHECRs in comparison with experimental data.
For this purpose CRPropa 3 was developed, with the physics processes based to a large extent on the original CRPropa 2.0~\cite{Kampert:2012fi}.
In this contribution we will summarize its main new features.

\section{Overview of CRPropa}
CRPropa is a publicly available simulation software for the Monte Carlo (MC) propagation of cosmic ray nuclei, photons and neutrinos through an extragalactic and galactic environment.
CRPropa simulates interactions such as electron pair-production, pion-production and photo-disintegration of nuclei with the diffuse extragalactic background radiation, as well as nuclear decay, and it includes secondary particles created in these interactions.

The simulations can be performed either in a one-dimensional (1D) or three-dimensional (3D) mode.
The 3D mode allows to define a 3D source distribution and takes into account the deflections of charged cosmic rays in extragalactic magnetic fields.
In the 1D mode the cosmological evolution of the sources and background radiation as well as the adiabatic energy loss of cosmic rays can be implemented.
CRPropa thereby enables the user to predict the spectra of UHECR (and of their secondaries), their composition and arrival directions.
All of these features are inherited by the new version CRPropa 3~\cite{Batista:2013gka}.

\section{New features of CRPropa 3}
CRPropa 3 features several significant advancements in the CRPropa development.
The main improvement is a new modular simulation layout, that allows for multiple simulation use cases, and for easier testing, maintenance and physics extensions.

\subsection{Code structure, steering and parallelization}
CRPropa 3 was completely rewritten with a new code structure, separating all aspects of the simulation into independent modules that correspond to individual photo-nuclear interactions, boundary conditions, observers etc.
The cosmic ray particle class serves as a single interface between the simulation modules.
The modules provide a method to update the cosmic ray particle according to the module's purpose.
The simulation itself is a user-defined sequence of simulation modules, that are called in turn to update the cosmic ray until the propagation is either completed or aborted.
Since there are no direct dependences between modules, any combination of modules can in principle be selected, allowing for multiple use cases and to study in detail individual propagation aspects.

Efficient MC propagation depends on dynamically adjusting the step size to accommodate for varying conditions, e.g.\ making smaller steps in regions of strong magnetic deflections.
A bidding system allows all modules of a simulation to bid for the next step.
The lowest bid is then selected as step size for the next iteration of the module sequence.
Therefore the propagation proceeds with the largest possible step that still ensures the numerical accuracy as defined by the user.

CRPropa 3 is written in C++ and interfaced to Python using SWIG~\footnote{\url{www.swig.org}}.
This allows to set up and steer simulations in a high level scripting language while all computations are performed with the underlying C++ code.
The SWIG interface enables cross-language polymorphism, which can be used to extend a CRPropa simulation directly from the Python script that runs it.
The user can for example write a custom simulation module in Python to be used in combination with the existing C++ modules.
While the Python usage is the advised steering mode, backwards-compatibility to the XML steering of CRPropa 2 is provided as well.

\begin{figure}[h]
  \begin{minipage}{0.4\textwidth}
    \includegraphics[width=0.9\textwidth]{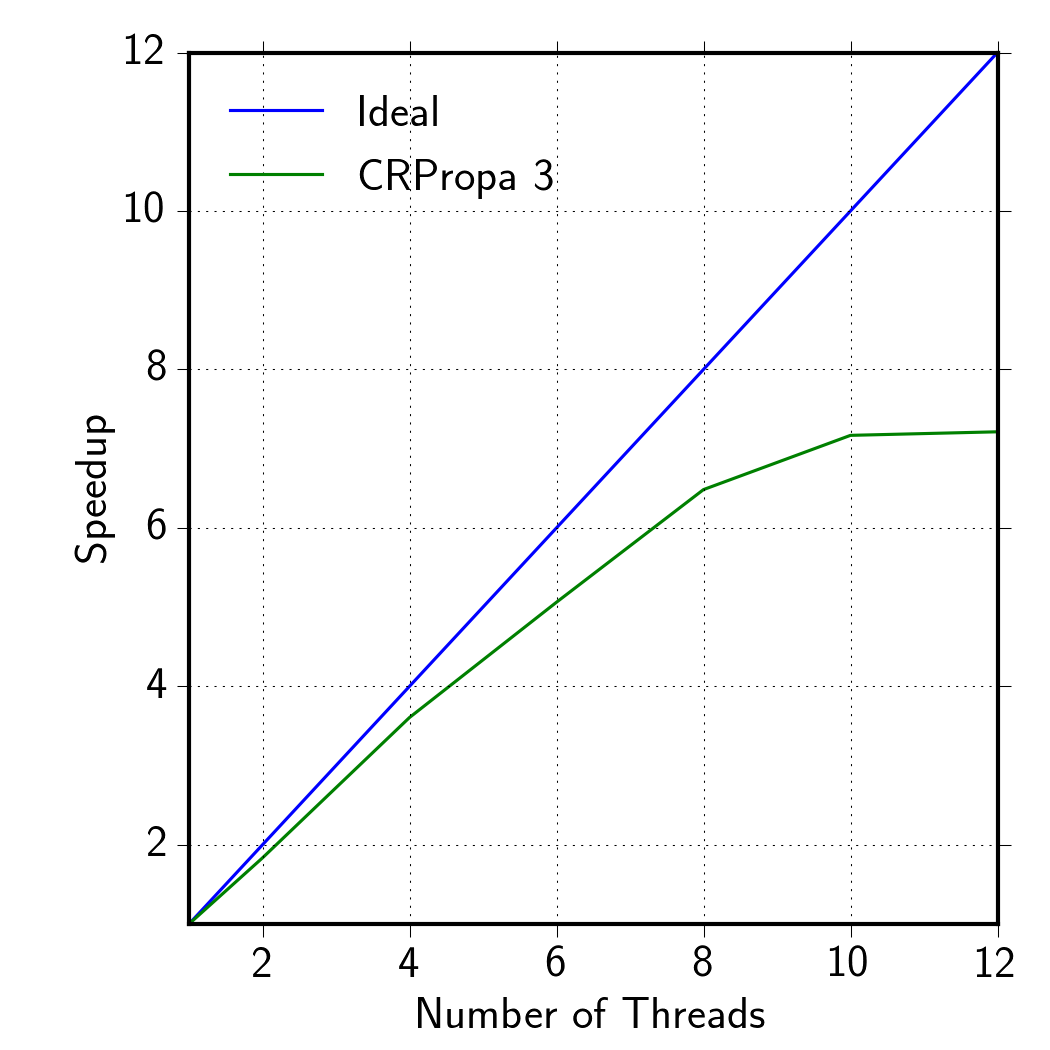}
    \caption{Speedup of CRPropa 3 in an example simulation of extragalactic propagation. The presence of non-parallelized sections limit the speedup to about 6-8 in typical simulations.}
    \label{fig:speedup}
  \end{minipage}
\hfill
  \begin{minipage}{0.55\textwidth}
    \centering
    \includegraphics[width=0.49\textwidth]{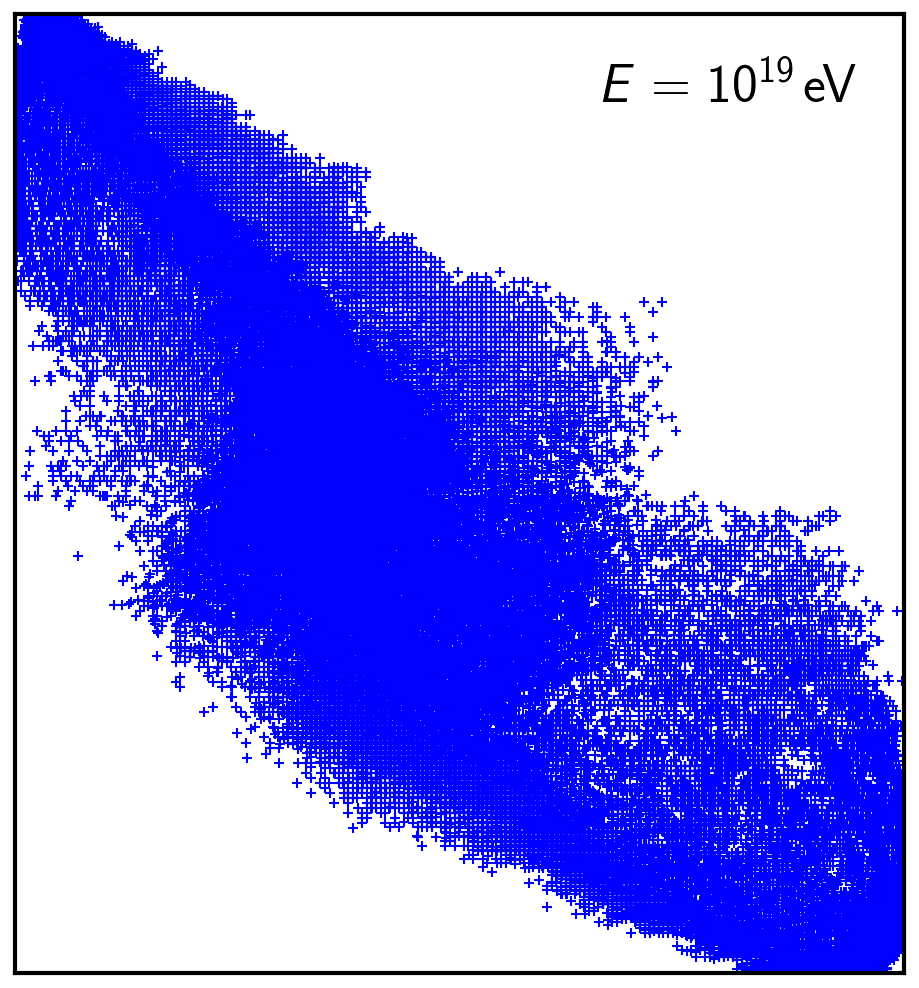}
    \includegraphics[width=0.49\textwidth]{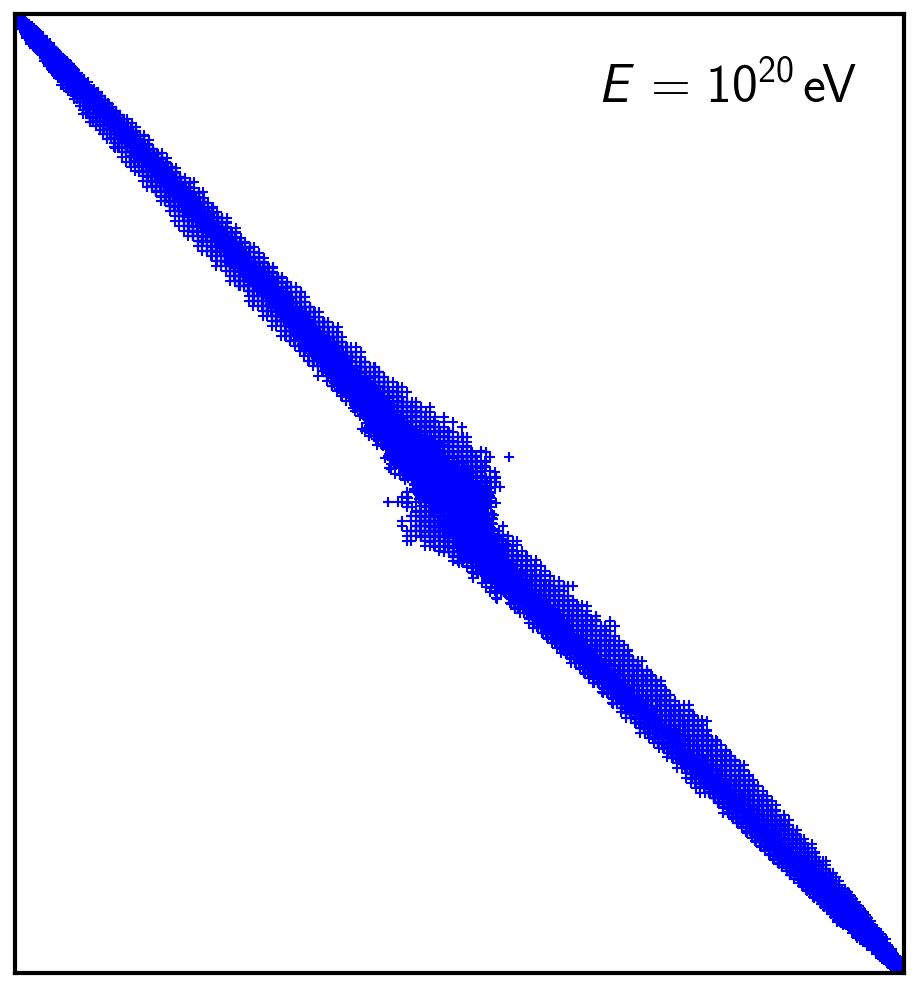}
    \caption{Representation of the non-zero matrix elements of the magnetic field lens for the JF12 model and two example energies, $10^{19}$ and $10^{20}$\,eV. Non-diagonal entries correspond to cosmic ray deflections, which are more pronounced with decreasing energy. The sparsity of the matrices is used to significantly lower the memory requirement.}
    \label{fig:matrix}
  \end{minipage}
\end{figure}

Cosmic ray propagation is a perfectly parallel task as interactions between cosmic rays are negligible.
Current multicore processors can therefore be adequately utilized by just running multiple simulation instances in parallel.
However, for better memory utilization, CRPropa 3 enables shared-memory multiprocessing using OpenMP~\footnote{\url{www.openmp.org}}.
This allows to use higher resolution magnetic fields and matter distributions in the simulation.
The parallelization occurs on the level of the module sequence with a dynamic distribution of cosmic rays among the available threads.
As the simulation modules are stateless, only a single instance of each module is needed.
The speedup is limited by the number of critical sections that are not thread safe and can only be executed by one thread at a time.
The critical section with the largest impact is the external library SOPHIA~\cite{Mucke:1999yb}, used to simulate photo-pion interactions. Thus, the speedup depends on the frequency of these interactions.
The speedup for a typical extragalactic simulation is shown in fig.~\ref{fig:speedup}.

\subsection{Magnetic field techniques}
\label{sec:magneticfieldtechniques}
Magneto-hydrodynamical simulations of structure formation provide models of the structured extragalactic magnetic field and matter distribution.
The challenge of these simulations is the large difference in scale between the simulation volume of $\sim (0.1 - 1\,\rm{Gpc})^3$ size and the required resolution to resolve the structures down to the level of galaxies $\sim 1 - 10 \rm{kpc}$.
Two of the techniques used in this context are smooth particles (SP) and adaptive mesh refinement (AMR).
In order to directly use the resulting extragalactic magnetic field models, CRPropa provides interfaces to the SP-code Gadget~\cite{Dolag:2008ya} and the AMR-code RAMSES~\cite{Teyssier:2001cp}.
The trade off, however, for the small memory demand of these techniques are large lookup times, which easily become the bottleneck for tracking cosmic rays.
In contrast, regular grids provide fast lookup times but easily require a terabyte memory.
As a compromise CRPropa offers the use of modulated grids, which are a combination of a small high-resolution vector grid that is periodically repeated to cover a larger volume, and a large low-resolution modulation grid that carries information about the large-scale structure.

\subsection{Galactic propagation}
Estimates for the strength of the galactic magnetic field imply that cosmic rays of rigidities $E/Z > 1$\,EeV, where $E$ is the energy and $Z$ the charge number, may be considerably deflected in the Galaxy, but without propagating diffusively.
Thus, due to the short galactic propagation as compared to extragalactic distances, energy loss processes are usually neglected and only magnetic deflections are considered.
CRPropa 3 enables the galactic propagation by providing models of the galactic magnetic field and making use of the modular simulation layout.
Both forward- and backward tracking are supported.
Forward tracking is computationally expensive as the Earth is a point-like target compared to galactic distances, resulting in a very small hit probability.
In the second approach, cosmic rays are propagated backwards from Earth to the galactic border.
However, it is not straightforward to connect galactic backtracking with extragalactic forward tracking.
As an alternative third approach, CRPropa 3 provides the lensing technique described and implemented in the PARSEC code~\cite{Bretz:2013oka}.
The lensing technique uses a set of matrices for different rigidities that are constructed from backtracking simulations with e.g. CRPropa 3.
The matrices transform cosmic ray directions at the galactic border into arrival directions observed at Earth.
The directions are binned with a HEALpix~\cite{Gorski:2004by} scheme into $\sim 50,000$ pixels for angular resolutions $< 1^\circ$.
The matrices thus have $\sim 50,000^2$ entries, nominally corresponding to 9\,GiB in single precision.
Sparse matrices are utilized to significantly lower the memory requirement to typically less than 10\,MB per matrix.
A representation of the non-zero elements of two example matrices is shown in fig.~\ref{fig:matrix}.
The lenses\footnote{Lenses for a number of galactic magnetic field models can be found on\,\newline\url{http://web.physik.rwth-aachen.de/Auger_MagneticFields/PARSEC/downloads.php}} can be used to transform entire arrival distributions through matrix-vector multiplications, or, in the context of individual cosmic rays, as a lookup table of precomputed trajectories.

Generic combinations of axisymmetric or bisymmetric spiral disc and halo fields can be considered as magnetic field models.
CRPropa 3 also implements the JF2012 model~\cite{Jansson:2012pc,Jansson:2012rt} including the random large-scale and turbulent small-scale component.
Additional user defined models can be easily implemented through Python extensions.

\subsection{4D simulations}
Cosmological effects such as the evolution of the background radiation are important when simulating the propagation of UHECRs.
In fact, including these effects in 3D simulations is necessary for anisotropy studies, except for the highest cosmic ray energies, where propagation distances are sufficiently short.
However, in the presence of magnetic deflections it is not possible to know {\it a priori} the effective propagation length and, therefore, the initial redshift of a cosmic ray that is observed at the present time.
Consequently, CRPropa 3 allows for 4D simulations, in which cosmic rays propagate both spatially and in time.
While computationally expensive, due to the additional loss of statistics in the time dimension, there are multiple applications for these simulations.
In addition to the aforementioned anisotropy studies, effects of magnetic suppression can be directly simulated for spectrum and composition studies.
4D simulations can also be an important tool for the validation of {\it a posteriori} corrections, such as the magnetic suppression parametrization in \cite{Mollerach:2013dza} and \cite{Batista:2014xza}.

\section{Example application}
In the following we present an example of a 3D simulation of cosmic ray nuclei through a structured universe and consider the galactic propagation with the lensing technique.
\begin{figure}[ht]
  \begin{minipage}{0.56\textwidth}
    \centering
    \includegraphics[width=0.9\textwidth]{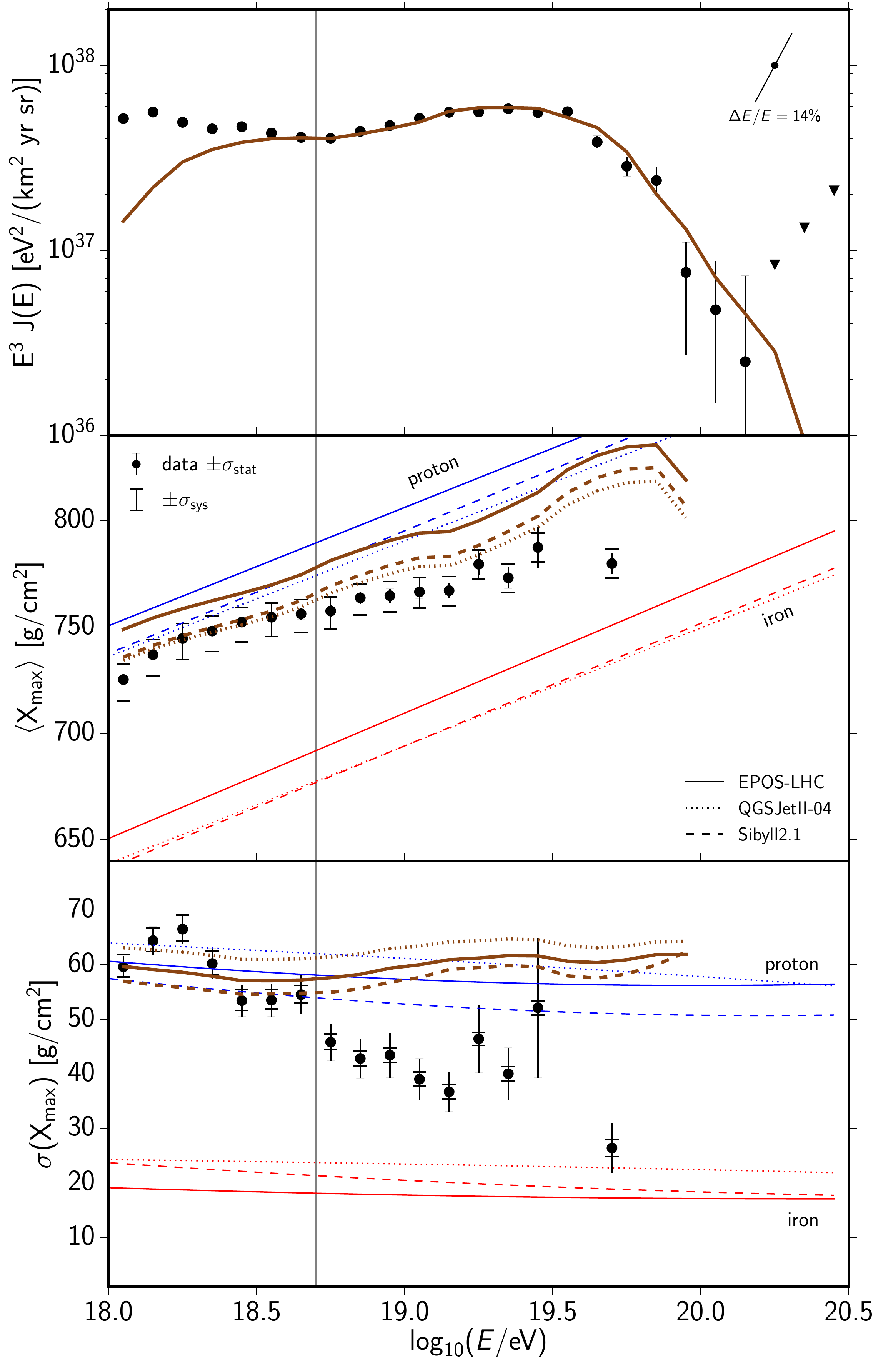}
    \caption{Energy spectrum (top) and moments of $X_\mathrm{max}$ (center and bottom) as measured by the Pierre Auger Collaboration \cite{ThePierreAuger:2013eja,Aab:2014kda} and predictions from the simulated scenario (thick brown lines). The red and blue lines show the predicted $X_\mathrm{max}$ moments for a pure composition of protons and iron, respectively. Solid, dotted and dashed lines show the cosmic ray shower predictions for different hadronic interaction models.}
    \label{fig:results1}
  \end{minipage}
  \hfill
  \begin{minipage}{0.4\textwidth}
    \includegraphics[width=\textwidth]{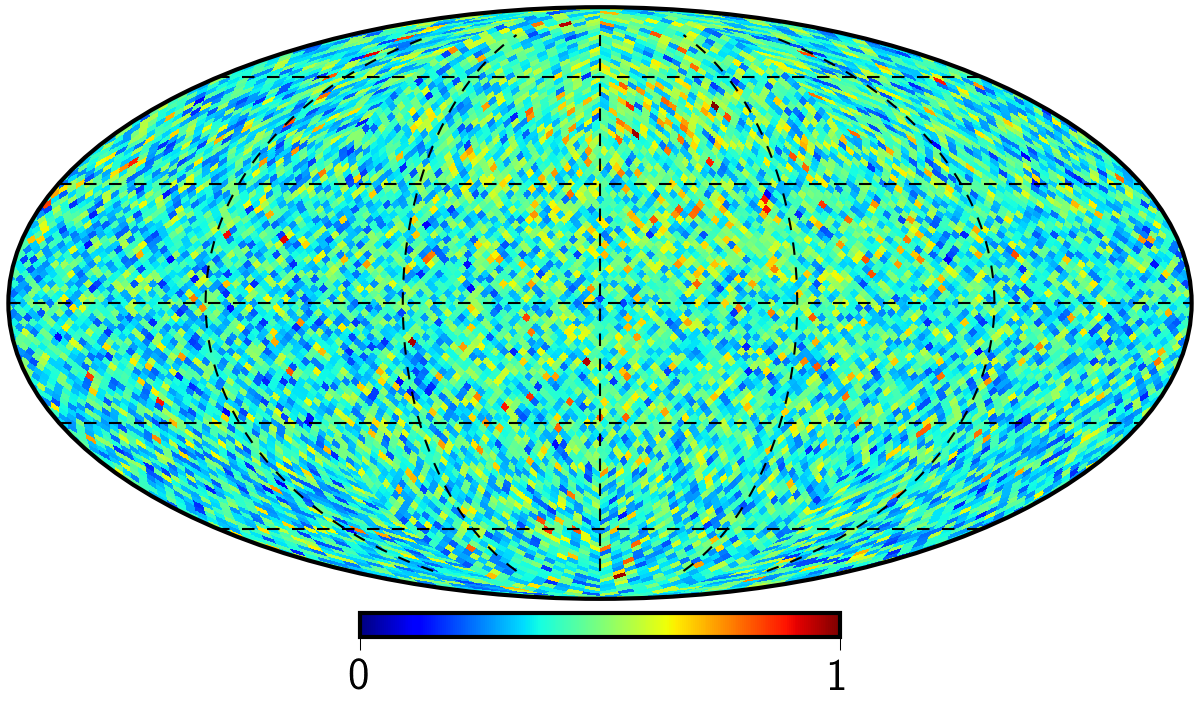}
    \includegraphics[width=\textwidth]{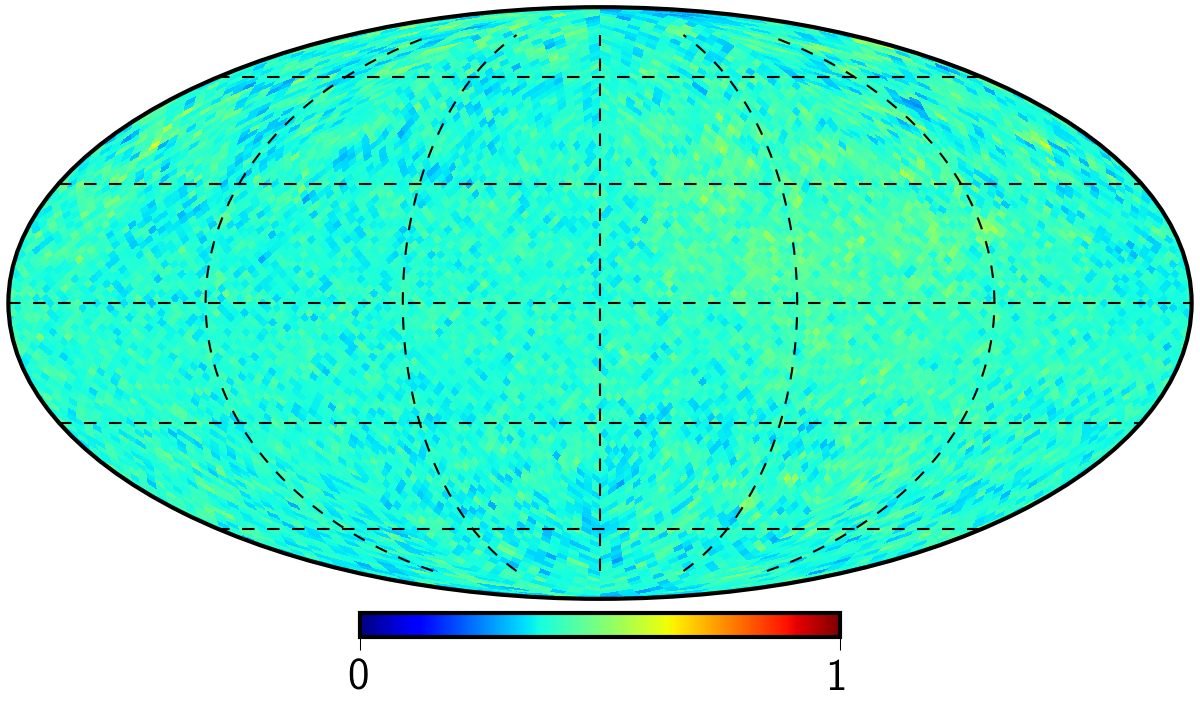}
    \caption{Distribution of arrival directions of events with energies \mbox{$E > 10^{18.7}$\,eV} before (top) and after galactic propagation (bottom). The events are binned in HEALpix maps and are normalized to the maximum event count of the map before galactic propagation.}
    \label{fig:results2}
  \end{minipage}
\end{figure}

As a model for the structured universe we use the simulated matter distribution and magnetic fields of Dolag {\it et al.}~\cite{Dolag:2004kp} and Miniati {\it et al.}~\cite{Sigl:2003ay}.
The Dolag simulation reproduces the local matter distribution by using a redshift survey as a constraint for the initial conditions.
The magnetic field arises from a primordial seed field, and is compatible with rotation measures of galaxy clusters.
However, the overall magnetic field strength is weaker compared to the Miniati simulation and leads to small UHECR deflections.
To emphasize the cosmic ray deflections, we make use of the relation of matter density vs. magnetic field strength obtained from the Miniati simulation. With this relation we translate the Dolag matter distribution into a distribution of the magnetic field strength and use this as a modulation field for a higher resolution turbulent field, as described in \ref{sec:magneticfieldtechniques}.
Both modulation and vector field are stored on a Cartesian grid. The modulation field covers a cubic volume of 132 Mpc edge length with a resolution of 300\,kpc. The vector field has a resolution of 50\,kpc and is initialized with a random turbulent realization of Kolmogorov power spectrum and a coherence length of 500\,kpc.

For the cosmic ray sources, we consider a continuous distribution that follows the matter density.
Using reflective boundary conditions for the cosmic rays, the contribution of sources up to 4\, Gpc distance is effectively taken into account.
The sources isotropically emit cosmic rays with a power-law spectrum and a rigidity dependent exponential cutoff.
The differential number of cosmic rays of charge number $Z_i$ and mass number $A_{i}$ is given by
\begin{equation}
\frac{{\rm d}N_i}{{\rm d}E} \propto x_i \, A_i^{1.8 - 1} \, E^{-1.8}\,\exp\left(- \frac{E}{Z_i \cdot 10^{19.8}\,\mathrm{eV}}\right)
\end{equation}
where $x_i$ is the relative abundance at equal energy per nucleon in absence of the cutoff.
The spectral index and cutoff rigidity are chosen as an example to approximately reproduce the observed spectrum above $10^{18.7}$\,eV.
As representatives for a mixed cosmic ray composition, the four isotopes hydrogen, helium-4, nitrogen-14 and iron-56 are selected with relative abundances $x_i =  1,\,0.5,\,0.3$ and 0.1, respectively.
The resulting energy spectrum as well as the mean and spread of the composition-sensitive $X_\mathrm{max}$ observable are shown in fig. \ref{fig:results1} in comparison with current measurements of the Pierre Auger Observatory \cite{ThePierreAuger:2013eja,Aab:2014kda}.
While a reasonable fit to the energy spectrum is achieved, the $X_\mathrm{max}$ moments are not well reproduced in this scenario.
The distribution of arrival directions of events with energies $E > 10^{18.7}$\,eV is shown in fig. \ref{fig:results2} both before and after galactic propagation, for which the full JF12 field model is considered.
While the small-scale anisotropy is seen to decrease through deflections in the galaxy, some anisotropic features remain.

\section{Conclusion}
In this contribution we have presented the public cosmic ray propagation code CRPropa 3.
We summarized the advantages of the new code structure,
the implementation of multi-processing and highlighted some of the new features, notably the galactic propagation, including the lensing technique, and the possibility to consider cosmological evolution in 4-dimensional simulations.
We applied CRPropa to a scenario in which a mixed composition of hydrogen, helium, nitrogen and iron was propagated through a combination of structured universe models, and considered deflections in the galactic magnetic field.
The simulation results were compared to measurements of energy spectrum and composition of the Pierre Auger Observatory.
The arrival directions were shown as well, to demonstrate the possibility of simulating all observables accessible to UHECR experiments.
More information on CRPropa can be found on https://crpropa.desy.de.

\section*{Acknowledgments}
This work was supported by the Deutsche Forschungsgemeinschaft (DFG) through the Collaborative Research Centre SFB 676 “Particles, Strings and the Early Universe”.
RAB acknowledges the support from the Forschungs- und Wissenschaftsstiftung Hamburg through the program “Astroparticle Physics with Multiple Messengers”.
GS was supported by the State of Hamburg, through the Collaborative Research program “Connecting Particles with the Cosmos” and by BMBF under grants 05A11GU1 and 05A14GU1.

\section*{References}
\begin{NoHyper}
\bibliography{bibliography}
\end{NoHyper}
\end{document}